\documentclass[12pt,preprint]{aastex}

\begin{document}

\title{GRB 060218: The nature of the optical-UV component}

\author{C.-I. Bj\"ornsson\altaffilmark{1}}
\altaffiltext{1}{Department of Astronomy, AlbaNova University Center, SE--106~91 Stockholm,
Sweden.}
\email{bjornsson@astro.su.se}

\begin{abstract}
The optical-UV component in GRB 060218 is assumed to be due to optically thick cyclotron emission. The key aspect of this model is the high temperature of the absorbing electrons. The heat input derives from nuclei accelerated in semi-relativistic internal shocks, like in ordinary gamma-ray bursts. Coulomb collisions transfer part of that  energy to electrons. Inverse Compton cooling on the X-ray photons leads to electron temperatures around $\sim 100$ keV. Such a high brightness temperature for the optical-UV emission implies an emitting area roughly equal to that of the thermal X-ray component. This suggests a model in which the radio, optical-UV and thermal X-ray emission are closely related: Although the optical-UV and thermal X-ray emission are two separate spectral components, it is argued that they both come from the photosphere of a quasi-spherical, continuous outflow, whose interaction with the circumstellar medium gives rise to the radio emission. The properties of GRB 060218, as measured in the co-moving frame, are similar to those of ordinary gamma-ray burst; i.e., the main difference is the much lower value of the bulk Lorentz factor in GRB 060218. The cyclotron absorption implies a magnetic field in rough equipartition with the matter energy density in the outflow. Hence, the magnetic field could have a dynamically important role, possibly with a magnetar as the central engine.
 \end{abstract}


\section{Introduction}
It was suggested early on that Gamma-Ray Bursts (GRBs) could be the result either of the core-collapse of massive stars or the merger of two compact objects in a binary system. The bi-modal distribution of duration times for GRBs observed by BATSE was taken as support for these ideas; long-duration GRBs were attributed to the former scenario while the latter scenario was thought to give rise to short-duration GRBs. The first direct evidence for such an origin for long-duration GRBs came from the association of SN 1998bw with GRB 980425 \citep[e.g.,][]{iwa03} and it was further strengthen by observations of GRB 030329/SN 2003dh \citep[e.g.,][]{sta03,hjo03}. With its highest sensitivity at frequencies lower than for BATSE, the BeppoSax satellite revealed the existence of bursts in which the energy output peaked in the X-ray range. \citet{hei01} argued that these X-Ray Flashes should be regarded as a separate class. Such a distinction has been questioned \citep[e.g.,][]{rho03,sak05}, since several of their properties are either similar to or constitute a continuous extension of those of BATSE-GRBs. The existence of classes or sub-classes of long-duration GRBs is still an open question. It has received further attention with the discovery of GRB 060505 and GRB 060614; for both of these, a possible associated supernova is weaker than SN 1998bw at least by a factor of one hundred \citep{fyn06,del06}. In addition, in spite of its long duration, GRB 060614 has several properties in common with short-duration GRBs. This has prompted \citet{geh06} and \citet{gal06} to argue for the need to consider formation scenarios for GRBs that go beyond the two originally proposed.

Several of the salient features of GRB 060218 conformed to those for a standard gamma-ray burst; for example, the prompt high-energy emission, although unusually long, exhibited the expected transition to an afterglow phase which also included a radio component.  Even on a more detailed level, GRB 060218 complied with several of the trends followed by the large majority of GRBs. Although the isotropic equivalent emitted energy ($E_{\rm iso}$), as well as the energy where it peaked ($E_{\rm p}$), were low \citep{cam06}, they obeyed the Amati relation \citep{ama02}, which was established for GRBs with considerably larger values for these quantities. \citet{lia06} showed that GRB 060218 also fits the luminosity-lag relation first found by \citet{nor00} for BATSE-GRBs. In contrast, the properties of the optical-UV component  set it aside from other GRBs; for example, the emission could be ascribed to a black-body component. There is also a black-body component apparent in the prompt X-ray emission. Indications of black-body emission in the prompt phase are not rare. In fact, it has been argued \citep[e.g.,][]{ryd06} that, in addition to the non-thermal emission, a black-body component is quite common in GRBs observed by BATSE. However, one of the interesting aspects of the thermal X-ray emission in GRB 060218 is that the implied low temperature makes a connection possible to the break-out of the shock in the associated supernova SN 2006aj \citep{cam06}. 

\citet{cob06} showed that GRB 060218 is similar to GRB 980425 and GRB 031203. They emphasized that even the detection by Swift of one such low luminosity GRB indicates a different origin as compared to classical GRBs. From spectral analyses of SN 2006aj, \citet{maz06} and \citet{mae07} inferred a mass of the progenitor star low enough for a neutron star to result from the core-collapse. This is in contrast to the belief that classical GRBs originate in stars massive enough for black holes to form. \citet{sod06} put forth similar arguments and suggested that the neutron star, in the form of a magnetar, acts as the central engine in GRB060218. They also pointed out that the radio observations strongly indicate a quasi-spherical outflow in GRB 060218 rather than the jet-like geometry normally invoked for GRBs. 

This makes it likely that in some fundamental aspect GRB 060218 differs from most other GRBs. Since the optical-UV component is a distinguishing feature of GRB 060218, the focus of the present paper is the nature of this component and how it relates to the other spectral components. The hope is that a consistent description will elucidate not only what makes GRB 060218 different but also why many of its properties are so similar to those of other GRBs. 

The paper is organized as follows. In \S~\ref{sect2}, a short summary is given of the various models proposed for GRB 060218. Some of their limitations are discussed and it is suggested that a consistent picture can be obtained by assuming that the optical-UV component is due to optically thick cyclotron emission. The details of such a model are presented in \S~\ref{sect3}. An application to GRB 060218 is done in \S~\ref{sect4} and its main physical characteristics are deduced. The implications of this analysis are discussed in \S~\ref{sect5} with a particular emphasis on the physical relation between the various spectral components.

\section{Modelling the emission from GRB 060218} \label{sect2}
\citet{cam06} discussed the implications of a scenario in which the thermal X-ray emission originates from behind the supernova shock \citep[see also][]{wax07}. They pointed out that the implied radiating surface is much smaller than expected from the emitting volume. In order to account for this discrepancy, they invoked a non-spherically expanding shock. Furthermore, the deduced emitting radius is much larger than the radius of the expected progenitor star, which implies a high mass-loss rate. The photosphere would then correspond to the radius where the wind becomes optically thin to electron scattering. After the shocked gas has become optically thin, the optical-UV emission could come from the expanding supernova ejecta. However, such an origin is not likely at earlier times and \citet{cam06} suggested that this part of the optical-UV emission was the low energy extension of the thermal X-ray component.
\citet{ghi07} emphasized that the required emitting surface of the optical-UV emission is much too large to be compatible with an extension of the thermal X-ray component and, instead, suggested that all of the optical-UV emission derived from a self-absorbed synchrotron source. Due to the high brightness temperature for such an emission mechanism, the emitting surface is small and, in analogy with the standard model for afterglows, they invoked a relativistic jet.

From the radio observations, \citet{sod06} deduced a quasi-spherical geometry for this spectral component and that it moves with a mildly relativistic velocity. It is worth emphasizing that, with the exception of the assumption of a synchrotron origin for the radio emission, these results are quite model-independent. Furthermore, \citet{sod06} argued for a mass loss rate of the progenitor star almost a factor $10^3$ smaller than that needed by \citet{cam06} in their model.
These rather disparate conclusions regarding the properties of the three spectral components make it hard to discern any physical relationship between them and, instead, suggest that they are quasi-independent. This is most apparent in the deduced geometries with a non-spherical shock for the thermal X-rays, a relativistic jet for the optical-UV emission and a mildly relativistic, quasi-spherical shock for the radio component.

As emphasized by \citet{cam06}, the key to interpreting the observations of GRB 060218 is the origin of the thermal components. It is intriguing that there are indications of similarities in the properties of the time variations of the thermal X-ray emission and the optical-UV emission; for example, in the X-ray component both the radius, $R$, and temperature, $T$, change marginally during the first $\sim 10^{4}$\,s, while $R^{2}T$ increases only by a factor $\sim2$\,-\,$3$ in the optical-UV component during the same time interval. The size of the emitting surface of the optical-UV component depends to some extent on the assumed reddening correction applied to the observed flux. As discussed by \citet{ghi07}, for plausible values of the extinction, the indicated size is almost a factor $10^3$ larger than that for the thermal X-ray component, assuming the same brightness temperature for both components. Motivated by the similarity of their time variations, an alternative interpretation is to assume that the emitting surfaces are the same but, instead, that their brightness temperatures differ. Since $kT\approx 0.17$\,keV for the thermal X-ray component, this implies a brightness temperature $kT\sim 100$\,keV for the optical-UV component. Just as for synchrotron self-absorption, in such a scenario, the flux in the optical-UV component  peaks roughly at the frequency where the absorption optical depth is unity and the brightness temperature is determined by the energy of the absorbing particles. The main difference is that a brightness temperature $kT\sim 100$\,keV implies  an emitting surface a factor $\sim 10^2$ larger than for synchrotron self-absorption ($kT \sim 10$\,MeV, neglecting relativistic effects). In the following section the basic properties are derived for a model in which the optical-UV emission is due to optically thick cyclotron radiation. 

\section{A cyclotron absorption model} \label{sect3}
The main properties of a optically thick cyclotron source are determined by the strength of the magnetic field, $B$, and the electron temperature. Following the standard assumption about the energy input for the prompt emission in GRBs, it will be assumed that internal shocks accelerate nuclei to a velocity $v_{\rm n}$ and that each nucleus experiences at least one shock during a dynamical time. The energy gained by the electrons in these shocks is radiated away as prompt emission in a time much shorter than the dynamical time-scale. The temperature of the electrons is instead determined by the accelerated nuclei, which transfer part of their energy to the electrons via Coulomb collisions. There could be other, collisionless mechanisms that transfer energy from the nuclei to the electrons, in particular in view of the expected rather strong magnetic field. The efficiency of such mechanisms are not-well-known and, therefore, hard to quantify. Hence, the transfer rate derived below should be regarded as a lower limit. The electron temperature is then determined by balancing this heating with cooling. Since it is assumed that the emission of both the thermal X-rays and optical-UV emission have the same emitting surface, the cooling of the electrons is due to inverse Compton scattering on the thermal X-ray photons. The resulting electron distribution is assumed to be Maxwellian. In analogy with the standard model for GRBs, the internal shocks occur in a semi-continuous outflow or "shell" moving with a bulk Lorentz factor $\Gamma$.

\subsection{Electron temperature} \label{sect3a}
The transfer of energy from nuclei to electrons depends on their relative velocity, $v$. A rough approximation is, therefore, $v=\rm{max}\left[v_{\rm n},v_{\rm e}\right]$, where $v_{\rm e}$ is the electron velocity. The transfer of momentum depends only on $v$ even in the relativistic limit. Relativistic effects enter only in the Coulomb logarithm, $\ln C$. Hence, for simplicity, relativistic effects in the co-moving frame are neglected.
In the co-moving frame a nucleus with mass $pm_{\rm p}$ and charge $ze$, where $m_{\rm p}$ is the proton mass, loses energy at a rate 
\begin{equation}
	\frac{{\rm d}E_{\rm n}}{{\rm d}t} = \frac{3}{2}\sigma_{\rm T}z^{2}n_{\rm e} m_{\rm e}c^{2}v
	\left(\frac{c}{v}\right)^{2}\ln C,
	\label{eq:1.1}
\end{equation}
where, $\sigma_{\rm T}$ is the Thomson cross section, $n_{\rm e}$ is the density of electrons and $m_{\rm e}$ is the electron mass. During one dynamical time $\left(\approx R/\Gamma c\right)$, a nucleus loses a fraction
\begin{equation}
	\frac{\Delta E_{\rm n}}{E_{\rm n}} = 3\tau_{\rm T}\frac{m_{\rm e}}{m_{\rm p}}\frac{z^2}{p}
	\frac{c}{v}\left(\frac{c}{v_{\rm n}}\right)^{2}\ln C
	\label{eq:1.2}
\end{equation}
of its energy, where $\tau_{\rm T}$ is the effective optical depth (i.e., within one horizon length) due to electron scattering.

An electron loses energy at an average rate
\begin{equation}
	\frac{{\rm d}E_{\rm e}}{{\rm d}t} = 4 \sigma_{\rm T}U_{\rm ph}c\frac{kT_{\rm e}}{m_{\rm e}c^2}
	\label{eq:1.3}	
\end{equation}
due to scattering on photons with an energy density $U_{\rm ph}$. Here, $T_{\rm e}$ is the electron temperature. Balancing heating and cooling (i.e., equating eqs. [\ref{eq:1.1}] and [\ref{eq:1.3}] taking into account that there are $z$ electrons for every nucleus) results in
\begin{equation}
	\frac{kT_{\rm e}}{m_{\rm e}c^2} = \frac{3}{8}\frac{m_{\rm e}}{m_{\rm p}}
	\frac{z^2}{p}\frac{c}{v}\frac{\ln C}{\eta},
	\label{eq:1.4}
\end{equation}
where $\eta = U_{\rm ph}/npm_{\rm p}$ is the photon energy density relative that in matter. Here, $n=n_{\rm e}/z$ is the density of nuclei.

There is a critical value of the electron temperature ($T_{\rm crit}$) corresponding to $v_{\rm n} = v_{\rm e} = v_{\rm crit}$. With the use of $kT_{\rm e} = m_{\rm e}v_{\rm e}^{2}/3$, one finds
\begin{eqnarray}
	\frac{kT_{\rm crit}}{m_{\rm e}c^2}& = &\left(\frac{\sqrt{3}}{8}\frac{m_{\rm e}}{m_{\rm p}}
	\frac{z^2}{p}\frac{\ln C}{\eta}\right)^{2/3}\nonumber\\ &=& \frac{1.8\times 10^{-2}}
	{\eta^{2/3}}\left(\frac{z^2}{p}\right)^{2/3}\left(\frac{\ln C}{20}\right)^{2/3}.
	\label{eq:1.5}
\end{eqnarray}
This can also be expressed as
\begin{equation}
	kT_{\rm crit} = \frac{9.1}{\eta^{2/3}}\left(\frac{z^2}{p}\right)^{2/3}\left(\frac{\ln C}
	{20}\right)^{2/3}   ~\rm keV
	\label{eq:1.6}
\end{equation}
or
\begin{equation}
	\frac{v_{\rm crit}}{c} =  \frac{2.3\times 10^{-1}}
	{\eta^{1/3}}\left(\frac{z^2}{p}\right)^{1/3}\left(\frac{\ln C}{20}\right)^{1/3}.
	\label{eq:1.7}
\end{equation}

There are two regimes for $T_{\rm e}$ corresponding to the relative values of $v_{\rm n}$ and $v_{\rm e}$. When $v_{\rm n}<v_{\rm e}$ (i.e., $v=v_{\rm e}$), it is seen from equation~(\ref{eq:1.4}) that $T_{\rm e} = T_{\rm crit}$ (implying $v_{\rm n}<v_{\rm crit}$), while $v_{\rm n}>v_{\rm e}$ gives $T_{\rm e} = T_{\rm crit}\left(v_{\rm crit}/v_{\rm n}\right)$ (implying $v_{\rm n}>v_{\rm crit}$).

It has been implicitly assumed that the nuclei lose only a small fraction of their energy during a dynamical time. However, cooling of the nuclei can be important. As seen from equations~(\ref{eq:1.2}) and (\ref{eq:1.7}), the requirement for the nuclei to cool in the regime  $v_{\rm n}>v_{\rm crit}$ is $\eta \tau_{\rm T}>0.37 p/z^{2}$. It is argued below that this range of parameters is not applicable for GRB 060218. Hence, for the case of interest here, cooling occurs for $v_{\rm n}<v_{\rm cool}<v_{\rm crit}$, where
\begin{eqnarray}
	\frac{v_{\rm cool}}{c} &=&  \left(3\tau_{\rm T}\frac{z^2}{p}\frac{m_{\rm e}}{m_{\rm p}}
	\frac{c}{v_{\rm crit}}\ln C\right)^{1/2}\nonumber\\ &=& 0.38~\tau_{\rm T}^{1/2}
	\left(\frac{z^2}{p}\right)^{1/2} \left(\frac{\ln C}{20}\right)^{1/3}\eta^{1/6}
	\label{eq:1.8}
\end{eqnarray}	
is the velocity below which all the injected energy is radiated away as thermal inverse-Compton scattered radiation. For $v_{\rm n}<v_{\rm cool}$, the electron temperature is likely to vary in the gas and would depend on the number of re-accelerations experienced by the nuclei during one dynamical time; for example, with only one acceleration the column density of electrons with $T_{\rm e} = T_{\rm crit}$ is smaller than that implied by $\tau_{\rm T}$ by a factor $\left(v_{\rm n}/v_{\rm cool}\right)^2$.

\begin{figure}
\plotone{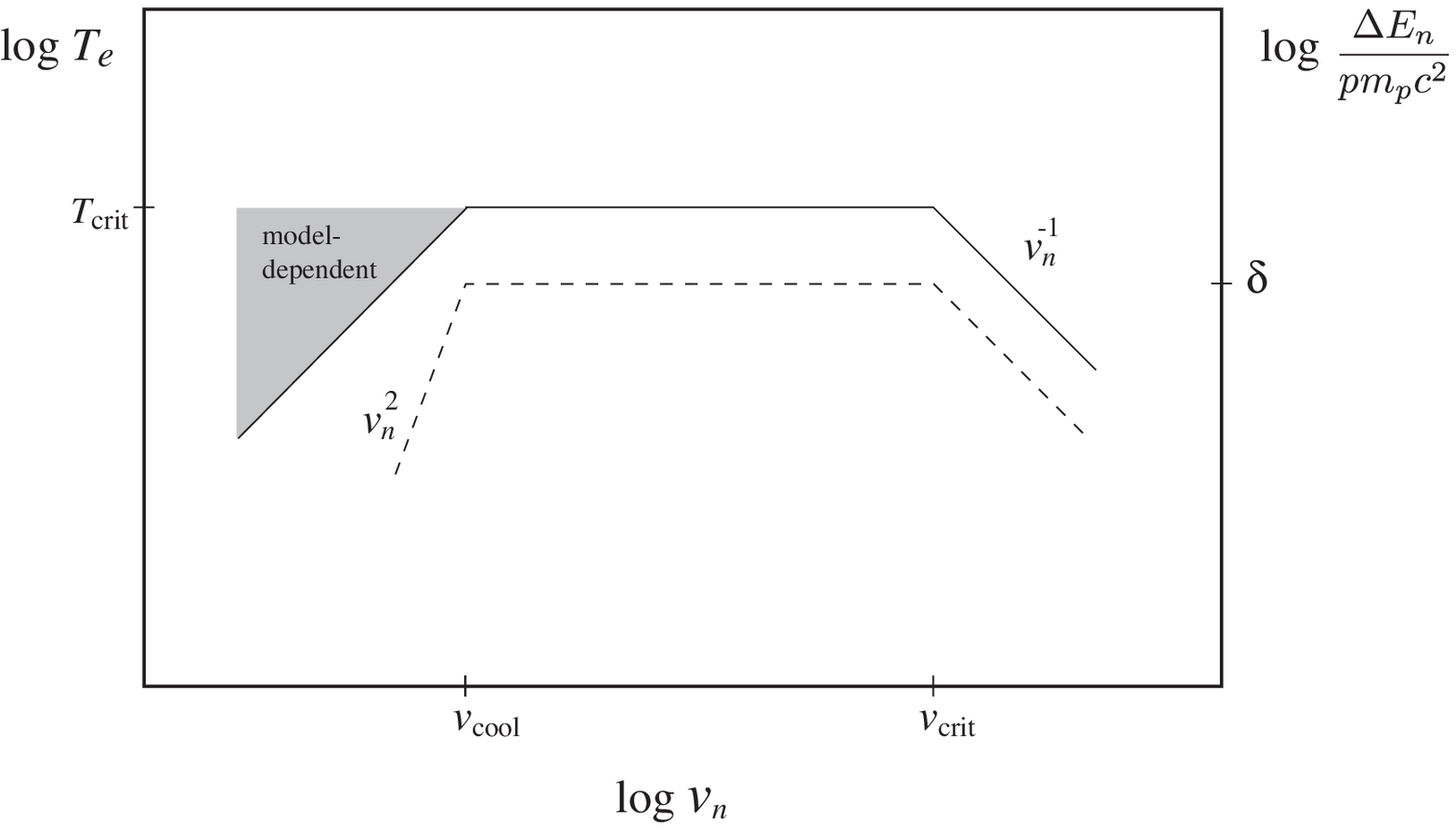}
\caption{A schematic representation of the variation of electron temperature ($T_{\rm e}$) with velocity of the nuclei ($v_{\rm n}$; {\it solid line}). For $v_{\rm n} < v_{\rm cool}$, the electron temperature is model dependent and the number of reaccelerations is important. Also shown is the energy transfered ($\Delta E_{\rm n}$) through Coulomb collisions from nuclei to electrons during one dynamical time  ({\it dashed line}) normalized to their rest-mass energy. The expressions for $T_{\rm crit}, v_{\rm crit}, v_{\rm cool}$, and $\delta$ are given in equations (\ref{eq:1.6}), (\ref{eq:1.7}), (\ref{eq:1.8}), and (\ref{eq:1.9}). \label{fig1}} 
\end{figure}

The energy radiated away per nucleus during one dynamical time is constant for $v_{\rm cool}<v_{\rm n}<v_{\rm crit}$ (i.e., $T_{\rm e} = T_{\rm crit}$). Normalized to its rest mass energy, it is given by $\Delta E_{\rm n}/pm_{\rm p}c^2 \equiv \delta$ (cf. eq. [\ref{eq:1.2}]), where
\begin{eqnarray}
	\delta &=& \frac{3}{2}\tau_{\rm T}\frac{z^2}{p}\frac{m_{\rm e}}{m_{\rm p}}\frac{c}{v_{\rm crit}}
	\ln C \nonumber\\ &=& 7.1\times 10^{-2}~\tau_{\rm T}\left(\frac{z^2}{p}\right)^{2/3}
	\left(\frac{\ln C}{20}\right)^{2/3}\eta^{1/3}.
	\label{eq:1.9}
\end{eqnarray}
Likewise, for $v_{\rm n}<v_{\rm cool}$
\begin{equation}
	\frac{\Delta E_{\rm n}}{pm_{\rm p}c^2} = \delta\left(\frac{v_{\rm n}}{v_{\rm cool}}\right)^2
	\label{eq:1.10}
\end{equation}
and for $v_{\rm n}>v_{\rm crit}$
\begin{equation}
	\frac{\Delta E_{\rm n}}{pm_{\rm p}c^2} = \delta\frac{v_{\rm crit}}{v_{\rm n}}.
	\label{eq:1.11}
\end{equation}
A schematic summary of the above results is given in Figure (\ref{fig1}).

\subsection{Relating kinetic and photon luminosities} \label{sect3b}
For a cold, relativistic flow moving with a bulk Lorentz factor $\Gamma$, the kinetic luminosity is given by
\begin{equation}
	L_{\rm kin} = 4\,\pi R^2 \Gamma^2 U_{\rm kin}c,
	\label{eq:2.1}
\end{equation}
where $U_{\rm kin}=npm_{\rm p}c^2$ is the energy density of matter as measured in the co-moving frame. Likewise, the photon luminosity can be written
\begin{equation}
	L_{\rm ph} = \frac{4}{3} \eta L_{\rm kin},
	\label{eq:2.2}
\end{equation}
where the numerical factor results from the assumption of an isotropic radiation field in the co-moving frame. With the use of the expression for the effective optical depth, $\tau_{\rm T} = \sigma_{\rm T} n_{\rm e} R/\Gamma$, one finds
\begin{equation}
	U_{\rm kin} = m_{\rm p}c^2\frac{p}{z}\frac{\tau_{\rm T} \Gamma}{\sigma_{\rm T} R}.
	\label{eq:2.3}
\end{equation}
Assuming a black-body spectrum for the photons and introducing $y = \Gamma h \nu_{\rm p}/m_{\rm e}c^2$, where $\nu_{\rm p}$ is the peak frequency of the photons in the co-moving frame, equations (\ref{eq:2.1}), (\ref{eq:2.2}) and (\ref{eq:2.3}) yield
\begin{equation}
	\Gamma = 4.5\times 10 \left(\frac{z}{p}\right)^{1/4} \frac{L_{\rm ph,45}^{1/8} y^{1/2}}
	{\tau_{\rm T}^{1/4}\eta^{1/4}}
	\label{eq:2.4}
\end{equation}
and
\begin{equation}
	R_{\rm 12} = 9.6\times 10^{-6} \left(\frac{z}{p}\right)^{1/4} \frac{L_{\rm ph,45}^{5/8}}
	{\tau_{\rm T}^{1/4}\eta^{1/4} y^{3/2}},
	\label{eq:2.5}
\end{equation}
where $L_{\rm ph,45} \equiv L_{\rm ph}/10^{45}$\,erg and $R_{\rm 12} \equiv R/10^{12}$\,cm.

\section{Application to GRB 060218} \label{sect4}
The supernovae associated with GRBs are generally classified as Type Ic indicating that neither hydrogen nor helium is present in the progenitor star. Although the presence of some helium in the outer layers is hard to exclude on observational grounds, it is likely that oxygen dominates. Therefore, in the following, $z=8$ and $p=16$ are used. As can be seen from the results derived above, the dependence on the chemical abundance is rather weak and, hence, the details of the composition of the progenitor star are not that important.

In the scenario envisaged in this paper, the optical-UV emission and the thermal X-rays are two different spectral components. However, for both of these components the observed radiation is emitted roughly where the scattering optical depth is unity (i.e., $\tau_{\rm T} \sim 1$) so that their emitting surfaces are approximately the same. As mentioned previously, this requires $kT_{\rm e} \sim 100$\,keV. It is seen from equation (\ref{eq:1.6}) that this implies $\eta \approx 0.10$ (i.e., $kT_{\rm e} = kT_{\rm crit} \approx 100$\,keV) and $v_{\rm n}/c$ in the range $\sim 0.5$\,-\,$0.8$. Hence, the internal shocks should be mildly relativistic. 

The temperature of the thermal X-ray component corresponds to $y \approx 1.3\times 10^{-3}$. Together with $L_{\rm ph} \approx 2.0 \times10^{46}$\,erg \citep{ghi07}, this gives from equations (\ref{eq:2.4}) and (\ref{eq:2.5}) $\Gamma \approx 1.7~\eta^{-1/4}$ and $R_{\rm 12} \approx 0.94~\eta^{-1/4}$. The value of $\eta \approx 0.10$ deduced above then implies $\Gamma \approx 3.0$ and $R_{\rm 12} \approx 1.7$. This shows that the out-flowing matter moves semi-relativistically and, in fact, the value of $\Gamma$ is close to the one derived by \citet{sod06} from the radio observations. The thermal X-ray component comprises about $20\%$ of the prompt emission. With $\eta \approx 0.10$, this implies that the total kinetic energy is $E_{\rm kin} \approx 2.0 \times E_{\rm iso}$; hence, $E_{\rm kin} \approx 1.2\times 10^{50}$\,erg.

With a Maxwellian distribution of electrons, the main factors 
determining the optical depth to cyclotron absorption are the 
electron temperature ($T_{\rm e}$) and the parameter \citep{C/L91}
\begin{eqnarray}
    \Lambda &\equiv& 4\pi\frac{e}{B}\frac{nR}{c}\nonumber\\
    &=&9.1\times 10^9\frac{\tau_{\rm T}}{B_{\rm 6}},
    \label{eq:2.6}
\end{eqnarray}
where $B$ is the magnetic field and $B_{\rm 6} \equiv B/10^6$\,G. This is, roughly, the optical depth at the cyclotron frequency, $\nu_{\rm B}\equiv eB/2\pi mc$. The frequency where the optical depth to cyclotron 
absorption is unity can then be written
\begin{equation}
    \nu_{\tau=1}\equiv q\frac{eB}{2\pi mc}.
    \label{eq:2.7}
\end{equation}
Values for $q$ appropriate for the present case have been calculated by 
\citet{Har88}. In this temperature range
$q$ increases somewhat more rapidly than linearly with $T_{\rm e}$, while the dependence on $\Lambda$ is much more modest \citep[cf.\,the parameter fitting to the 
numerical results in the non-relativistic limit done by][]{C/L91}. At 
$kT_{\rm e}=100$\,keV and $\Lambda\sim10^{10}$\,-\,$10^{11}$, $q\approx10^2$ in the ordinary 
plasma mode and somewhat larger in the extraordinary mode \citep{Har88}. One may note that with $B_{\rm 6} \sim 1$, a self-absorbed synchrotron source has $\gamma_{\rm abs} \sim 20$, where $\gamma_{\rm abs}$ is the Lorentz factor of the relativistic electrons emitting at the self-absorption frequency. Hence, in the present situation, $q$ is only a factor of a few smaller than $\gamma_{\rm abs}^2$ so that the deduced value of $B$ is roughly the same as that needed for a self-absorbed synchrotron source.

The UVOT light curves reach a maximum roughly at the same time, indicating that it is due to a maximum in $R^2 T_{\rm e}$. However, after this maximum the light curves decline more rapidly for the higher frequencies, which is due to the peak frequency entering the UVOT energy range. The initial rise of the flux is consistent with being due to an increasing value of $R$ as measured from the thermal X-ray component; hence, $T_{\rm e} \approx constant$ during this phase. If this is true also for the later phase of increasing flux, it is plausible that both the decline of the flux and the decrease of the peak frequency are due mainly to a decreasing value of $T_{\rm e}$. Since the peak frequency enters the UVOT energy range soon after maximum flux, it is then likely that the peak frequency also during the phase of increasing flux was not much larger than the value measured during the decline phase. 

The peak frequency of the flux in the optically thick-to-thin transition is normally somewhat larger than where the optical depth is unity. Together with $\Gamma \approx 3.0$ and the value of $q$ discussed above, this shows that the observed peak frequency is roughly a factor $10^3$ larger than $\nu_{\rm B}$ as measured in the co-moving frame. The observed peak frequency is approximately $2\times 10^{15}$\,Hz, which implies $B_{\rm 6} \approx 0.6$. This corresponds to an energy density $U_{\rm B}\sim 1\times10^{10}$\,erg/cm$^3$. The energy density in matter is obtained from equation (\ref{eq:2.3}) as $U_{\rm kin} \approx 7.9\times 10^9$\,erg/cm$^3$. Hence, $U_{\rm B} \sim U_{\rm kin}$, indicating that the magnetic field could have a dynamically important role.

The maximum of the optical-UV emission occurs later than that for the thermal X-ray component. The decline of the prompt GRB-phase is likely due to a decreasing efficiency of the internal shocks and, in the present model, accompanied by a decreasing value of the optical depth, $\tau_{\rm T}$. As is seen from equation (\ref{eq:1.8}), this actually increases the range of shock velocities over which $T_{\rm e} = T_{\rm crit}$. The value of $\eta$ and, hence, $T_{\rm crit}$ (see eq. [\ref{eq:1.6}])  depends on the details of the model. However, since the main factor determining the optical-UV emission is $T_{\rm e}$, a gradual decrease of the shock efficiency does not necessarily imply a similar decrease in this component. In fact, the decline could be delayed until the energy released by internal shocks becomes smaller than that emitted by the optical-UV component assuming the above value of $T_{\rm e}$. The present model is then no longer applicable and the value of $T_{\rm e}$ would drop.

The semi-continuous outflow together with the large cyclotron optical depths imply that emission at smaller frequencies come from somewhat larger radii in the optical-UV component. As a result, the source structure is likely to be inhomogeneous and, hence, the spectral flux should increase less rapidly than $\nu^2$. Therefore, the extinction towards GRB 060218 cannot be inferred by assuming a $\nu^2$-spectrum. The low value of the extinction suggested by \citet{sol06} is compatible with the expectations from the present model.

The total energy of scattered radiation normalized to that in seed photons is, roughly, $4\tau_{\rm T}kT_{\rm e}/m_{\rm e}c^2$ (i.e. $\delta/\eta $). With $\tau_{\rm T} \sim 1$ and $kT_{\rm e} \sim 100$\,keV, this is of order unity and, hence, the emerging spectrum will be substantially affected. The details of the spectrum is model dependent, since the scattering process is both time dependent and occurs at $\tau_{\rm T} \sim 1$. This is a non-trivial problem and \citet{ree05} have discussed some of the characteristic spectral features expected from the more general case of internal energy dissipation at optical depths $\tau_{\rm T} \gtrsim 1$. A related question concerns the connection between the thermal and the non-thermal X-ray components. It has been assumed above that the Coulomb heated electrons cool on the thermal X-ray photons only. However, it is possible that the non-thermal component is produced close enough to the thermal component for it to contribute to the cooling of the electrons. Although this would effect the spectrum of the scattered radiation, the conclusions regarding the optical-UV component would not change, except that the required total kinetic energy need to be increased proportionately in order to keep the value of $\eta$ roughly constant.

Although the scattered radiation is likely to contribute substantially to the spectral distribution all the way up to the electron temperature (i.e., $\sim 100$\,keV) during the prompt phase, its distinct signature is harder to establish. The best phase to look for clear signs of scattered radiation may be during the decline of the prompt emission, since the value of $\tau_{\rm T}$ is then expected to decrease. It is interesting to note that the X-ray spectra shown by \citet{ghi07} during this phase contain a power-law component, whose flux decreases with time while the spectral index, $\alpha$, increases (here, $F_{\nu}\propto \nu^{-\alpha}$ is the spectral flux). This is consistent with a scattering origin for this component in a situation where  $\tau_{\rm T}<1$ and decreases with time. From the light curve presented by \citet{cam06} it is likely that this power-law component is the afterglow of GRB 060218. Such an origin for the afterglow would suggest a more direct connection, than is normally thought to be the case, between the prompt and afterglow components. However, the afterglow of GRB 060218 has a few characteristics that distinguish it from other afterglows. Although the temporal behavior of the afterglow is consistent with those of other long GRBs, its spectral index is much steeper than usual; for example, \citet{ghi07} measure $\alpha\approx 3.6$ at $ 40,000$\,s (the afterglow starts to dominate the X-ray flux at approximately $10,000$\,s.). Furthermore, \citet{sod06} finds $\alpha\approx 2.2$ five days after the burst, which suggests that after an initial softening the afterglow spectrum hardens with time. Since the total X-ray flux is still dominated by the soft emission, an additional component could have appeared at higher frequencies without affecting the light curve. A possible candidate is contribution from the cooling electrons behind the forward shock. A synchrotron self-absorption frequency of approximately $4$\,GHz \citep{sod06} on day five, leads to a synchrotron cooling frequency $\nu_{\rm c}\sim 10^{16}$\,Hz, assuming equipartition between relativistic electrons and magnetic fields. The spectral index below $\nu_{\rm c}$ implies $\nu F_{\nu} \sim constant$ ($\sim 10^{-14}$\,erg/s\,cm$^2$) above $\nu_{\rm c}$. This value is close to that observed at $\nu \approx 10^{18}$\,Hz by \citet{sod06}.

\section{Discussion} \label{sect5}
Different scenarios have been suggested for GRB 060218 and its afterglow depending on the observed frequency range considered. As mentioned in \S~\ref{sect2}, the physical properties derived from these partial models are such that it is hard to regard them as different aspects of one overall model;  in some cases, the conclusions drawn even contradict each other (e.g., geometry of the afterglow and mass loss rate of the progenitor star). The key aspect of the modeling is the origin of the optical-UV emission. In this paper it is suggested that this emission derives from optically thick cyclotron radiation. The absorbing electrons are heated through Coulomb interaction with nuclei accelerated in the same internal shocks which give rise to the prompt emission. It is shown in \S~\ref{sect4} that this allows a consistent description of the observed properties. The thermal X-ray component and the optical-UV emission are both emitted from the photosphere of a continuous outflow, while the radio emission originates from behind the forward shock caused by the interaction of this outflow with the wind from the progenitor star. The velocity deduced from the radio observations is roughly the same as that associated with the optical-UV and X-ray components. Hence, except for the velocity differences giving rise to the internal shocks, the bulk velocity of  the outflow does not vary much during the time of photospheric emission.

The main difference between GRB 060218 and most of the other long duration GRBs is the low value deduced for the bulk Lorentz factor of its outflow; the flow is only mildly relativistic with $\Gamma \approx 3$. This value is a factor of a few hundred smaller than that normally thought appropriate for long duration GRBs. This makes it possible to attribute many of the observed differences between GRB 060218 and other long duration GRBs as due mainly to different values of $\Gamma$, i.e., the properties of GRB 060218 in the co-moving frame do not differ much from those of other long duration GRBs. These properties include both the matter density and the thickness of the region giving rise to the prompt phase, which implies that the isotropically equivalent ejected mass associated with the burst is also the same as that for other GRBs. It has been suggested \citep[e.g.,][]{ryd06} that in many GRBs observed by BATSE, there is, in addition to the non-thermal emission, also a thermal component present. As measured in the co-moving frame, the typical temperature of this thermal component is close to that observed in GRB 060218. Furthermore, the ratio of thermal to non-thermal emission in GRB 060218 lies within the range deduced for the GRBs observed by BATSE. 

The emission from normal long-duration GRBs is thought to come from a narrow jet. An associated supernova is usually assumed to be a separate component in the sense that its emission comes from the remaining, major part of the progenitor star. Its properties are then not directly affected by the jet and, hence, they should conform to those of standard Type Ic supernovae, although there could be some observed features characteristic for the particular subgroup giving rise to GRBs. \citet{li07} and \citet{ghi07} emphasized that it is hard to associate either the optical-UV or the thermal X-ray emission in GRB 060218 with the breakout of a standard supernova shock. As discussed in \S~\ref{sect4}, in the present scenario, the geometry of the outflow as deduced from the optical-UV and thermal X-rays is quasi-spherical, which is in line with the robust conclusions drawn from the radio observations \citep{sod06}.
This suggests that both the radio emission from the forward shock and the photospheric emission giving rise to the optical-UV and X-ray emission are part of the same quasi-spherical outflow. Hence, there is no real difference between GRB 060218 and the associated supernova and, instead, they should be regarded as two aspects of the same quasi-spherical outflow. The main differences with a standard Type Ic supernova are the semi-relativistic shock velocity and the large amount of energy carried by this high velocity matter.

The mass loss rate  of the progenitor star deduced from observations differs considerably. \citet{cam06} argued for a high value ($\gtrsim 10^{-4} M_{\odot}\,\rm yr^{-1}$), while \citet{sod06} claimed a low value ($2\times 10^{-7} M_{\odot}\,\rm yr^{-1}$), both with an assumed wind velocity of $10^3\, \rm km\, s^{-1}$. It was shown in \S~\ref{sect4} that in order to account for the optical-UV component as well as the thermal X-ray emission, the kinetic energy associated with the outburst needs to be roughly a factor of $2$ larger than that emitted as photons. This is about a factor $100$ larger than the kinetic energy deduced from the radio observations by \citet{sod06} assuming a standard continuous outflow from a core-collapse supernova explosion. Since the value of the Lorentz factor of the outflow derived in \S~\ref {sect4} is close to that deduced by \citet{sod06}, the associated kinetic energies should be similar. However, it should be noted that the kinetic energy directly obtained from the radio observations is that pertaining to magnetic field and relativistic electrons. Hence, an estimate of the total kinetic energy  necessitates assumptions about the fraction of thermal energy behind the forward shock going into magnetic field ($\epsilon_{\rm B}$) and relativistic electrons ($\epsilon_{\rm e}$). \citet{sod06} used $\epsilon_{\rm B} = \epsilon_{\rm e} = 0.1$. Values of $\epsilon_{\rm B}$ and $\epsilon_{\rm e}$ a factor of $\sim 100$ smaller than this are needed in order to bring the value of the total kinetic energy in line with that derived in \S~\ref{sect4}. Such a lowering of the values for $\epsilon_{\rm B}$ and $\epsilon_{\rm e}$ would also cause the inferred mass loss rate of the progenitor star to increase to a few times $10^{-5} M_{\odot}\,\rm yr^{-1}$, which is in the range normally observed for Wolf-Rayet stars. Another possibility is that the momentum input to the forward shock is not continuous, like in ordinary supernovae, but instead is quasi-instantaneous with a shell-like structure as in the canonical fireball model for GRBs. For the latter case, a lower mass loss rate for the progenitor star is possible, since the deceleration phase would not have started at the time of the radio observations.

Canonical long-duration GRBs are normally thought to be associated with the core-collapse of the most massive stars, in which a black hole is formed rather than a neutron star. However, spectra of the supernova SN 2006aj, which accompanied GRB 060218,  both in its early \citep{maz06} and late phases \citep{mae07} indicate the explosion of a rather low mass star and, hence, the formation of a neutron star. This has lead \citet{maz06} to suggest that GRB 060218 is powered by a magnetar. A similar suggestion has been made by \citet{sod06} in their analysis of the non-thermal radiation from GRB 060218. It was shown in \S~\ref{sect4} that ascribing the optical-UV component to optically thick cyclotron radiation implies an energy density in magnetic fields which is comparable to that in matter. Although the conversion of magnetic energy into kinetic energy at large distances from a rapidly rotating magnetar is not well understood \cite[e.g.,][]{buc07}, the unusual properties of the early optical-UV emission from GRB 060218 could be direct evidence for a dynamically important role of a magnetar.

\acknowledgements

This research was supported by a grant from the Swedish Natural 
Science Research Council. Assistance from Luis Borgonovo is gratefully acknowledged.

\clearpage

\end{document}